# Rate-Equation Modelling and Ensemble Approach to Extraction of Parameters for Viral Infection-Induced Cell Apoptosis and Necrosis

Sergii Domanskyi,[a]  Joshua E. Schilling,[a]  Vyacheslav Gorshkov,[b]
Sergiy Libert,[c]*  Vladimir Privman,[a]**

[a]Department of Physics, Clarkson University, Potsdam, NY 13676
[b]National Technical University of Ukraine — KPI, Kiev 03056, Ukraine
[c]Department of Biomedical Sciences, Cornell University, Ithaca, NY 14853
*libert@cornell.edu
**privman@clarkson.edu

**ABSTRACT:** We develop a theoretical approach that uses physiochemical kinetics modelling to describe cell population dynamics upon progression of viral infection in cell culture, which results in cell apoptosis (programmed cell death) and necrosis (direct cell death). Several model parameters necessary for computer simulation were determined by reviewing and analyzing available published experimental data. By comparing experimental data to computer modelling results, we identify the parameters that are the most sensitive to the measured system properties and allow for the best data fitting. Our model allows extraction of parameters from experimental data and also has predictive power. Using the model we describe interesting time-dependent quantities that were not directly measured in the experiment, and identify correlations among the fitted parameter values. Numerical simulation of viral infection progression is done by a rate-equation approach resulting in a system of "stiff" equations, which are solved by using a novel variant of the stochastic ensemble modelling approach. The latter was originally developed for coupled chemical reactions.

**KEYWORDS:** rate equation; parameter extraction; viral infection; apoptosis; necrosis

# 1. INTRODUCTION

Recently, we developed a novel modelling approach[1,2] to describe the kinetics of cellular processes and cell-clustering and connectivity, using percolation theory.[3-7] Such modelling is based on ideas of statistical mechanics and yields results of interest to fields of aging and longevity. It can qualitatively reproduce certain experimentally observed features related to tissue "viability" and integrity (connectivity). We used similar physiochemical-kinetics/statistical mechanics approaches to suggest strategies for the development of materials with self-healing and self-damaging properties.[8-11] In the case of cell-population kinetics studies, for cell cultures cellular dynamics involves description of the rates of various processes, such as cell division, senescence, apoptosis/cell death, etc. In addition to global processes, we can also consider local cross-influences of the various cell types on the rates of these processes, for example, the influence of senescent cells on the replication of neighboring cells.

One of the most interesting topics of research is the ability of cellular systems to resist various "stresses." From the modelling point of view, dense two-dimensional structures are the most suitable for yielding detailed data of the time-dependence of both the cell numbers and their spatial cluster structure for various cell types, in response to various environmental insults. In this context, accurate experimental data can be collected using classical tissue culture techniques, and we have an ongoing experimental program that is expected to yield detailed cellular dynamics of dermal fibroblasts collected from various canine breeds. In addition to percolation modelling, one can use mean-field type rate equations to investigate cellular dynamics in more detail. Mean-field rate equations have been used successfully in describing similar processes in the context of self-healing materials.[8-10]

Indeed, in many situations, including dense clusters and low-confluency (dilute) systems, the rate-equation approach can be used successfully, without the need to address spatial fluctuations in connectivity.[4,12-14] Detailed time-dependent data for the dynamics of dense two-dimensional layers of fibroblast cells subject to "stresses" are of interest in the studies of aging (chemical or physical insults), but are not available thus far. However, interestingly, there are tabulated data[15] for the dynamics of this type of cells in a low-confluency culture upon viral infection.

In this work we consider the data of Ref. 15 for the time-dependence of the fraction of various cell counts — healthy, apoptotic, and necrotic. In these experiments,[15] the authors infected cells with a known amount of viral particles and monitored cell health in real time by observing the condition of cellular membrane and DNA fragmentation. Cellular membrane was visualized using Acridine Orange staining and nuclear DNA was visualized with Ethidium Bromide. By counting cells with intact membranes, "blebbing" membranes, permeabilized membranes, and fragmented nuclear DNA, the authors tracked[15] the fractions of cells dying via apoptosis (programmed cell death, as further described below) or necrosis (direct cell death due to damage, here by infection, etc.) in real time.

Here, we develop a model, using rate equations that are typically used to model different chemical and biochemical processes,[16-20] to describe cellular dynamics during viral infection. Rate equations offer an average or mean field approximation of dynamics of the system, which is applicable to the low-confluency cell culture considered here. It was found, however, that solving the set of rate equations numerically with the conventional Runge-Kutta 4$^{th}$ order method[21] would require a large computational effort due to having large and small rates mixed: the so-called "stiff" equations.[22] For example, we found that the rates of cell necrosis and of virus replication in the infected cells differ by six orders of magnitude at certain time scales. The adaptive-step Runge-Kutta-Fehlberg[23] method, RKF45,[24] is somewhat better but still numerically prohibitive. Therefore, we utilized an "ensemble" approach, based on stochastically evolving a large sample of objects (cells) labelled with various cellular properties (healthy, infected to various degrees, undergoing processes of apoptosis and necrosis, etc.) and changing in time. This approach has proved computationally efficient, requires averaging over only a few realizations to accurately describe the dynamics of our rate-equation system, as detailed in the Appendix, and therefore, can be of interest in other situations involving similar systems of rate equations.

The model is set up in Sec. 2, in which we describe the parameters required to describe all the considered kinetic processes. Not surprisingly, this type of modelling requires more than a few parameters, and therefore the data[15] do not determine all of them independently, with good precision. Thus, some of the parameters were taken as typical values available in the literature. A good number of parameters, to which the data are particularly sensitive, could be determined with high accuracy. These results are described in Sec. 2 and 3. Section 3 is devoted to results and discussion. Generally, a consistent picture of the system dynamics, with reasonable (as compared to results in the literature for this and other systems) process rate estimates is obtained, allowing us to extract some new parameter values that have not been measured experimentally.

## 2. MODEL DESCRIPTION

In order to model the processes in the considered experimental system,[15] we have to make certain assumptions on the cell types to consider, as well as on their dynamics. The concentration of the live cells will be denoted by $C_i(t)$, where the subscript $i = 0,1,2,...$ stands for the number of the viral genomes in the cell. This is a rather simplistic description of the "degree of infection" of the cell, convenient in the present modelling context. Here $t$ denotes the time, and $C_0(t)$ is then the number of the healthy non-infected cells. Initially, we have a low-confluency culture of such cells. The concentration of viruses will be denoted by $V(t)$, where initially for this experiment[15] the multiplicity of infection, termed MOI, was

$$V(0)/C_0(0) = 1. \tag{1}$$

As cells get infected some of them will initiate apoptosis and eventually die. We denote the respective concentrations by $A(t)$, and $D(t)$; see the Nomenclature Chart, Table 1. We denote by $N(t)$ the concentration of cells dead by necrosis. The total concentration of all the cell types is

$$S(t) = \sum_{i=0,1,2,...} C_i(t) + A(t) + D(t) + N(t), \tag{2}$$

and the data tabulated by Kumar et al.[15] were for the fractions of $(A + D)/S$, $(D + N)/S$, and $N/S$, for several times, $t$, during the experiment. Our data fit obtained with the preferred parameter set is shown in Fig. 1. As described below, some of the model parameters (introduced shortly) are well fitted based on the present data, some were estimated based on the literature results for other related systems, and some parameters are actually correlated with each other (the data do not fully determine them).

| | |
|---|---|
| | **Table 1.** Nomenclature Chart. |
| $C_0(t)$ | concentration of viable, non-infected cells |
| $C_{i=1,2,\ldots}(t)$ | concentration of non-apoptotic infected cells containing $i$ viral genomes |
| $V(t)$ | concentration of the viruses outside of the cells |
| $A(t)$ | concentration of apoptotic cells |
| $D(t)$ | concentration of cells dead due to apoptosis |
| $N(t)$ | concentration of cells dead due to necrosis |
| $S(t)$ | total concentration of cells |
| $n$ | number of viral genomes inside a cell beyond which the rate of necrosis saturates |
| $m$ | number of the viral genomes beyond which the cell is considered "well-infected" |
| $k$ | number of viral genomes beyond which the rate of their production saturates |
| $R$ | rate of cell division |
| $I_{i=0,1,\ldots}(t)$ | rate for infection: penetration of viruses into a viable or already infected cell |
| $r$ | constant in the rate of infection |
| $P_{i=1,2,\ldots}$ | rate of virus production inside viable non-apoptotic cells |
| $p$ | constant in the rate of virus production |
| $B_{i=1,2,\ldots}$ | rate at which viruses exit a cell with $i$ viral genomes in it |
| $b$ | constant in the rate of viruses exiting cells |
| $Q_{i=1,2,\ldots}$ | rate at which infected cells respond to the infection by initiating apoptosis |
| $q$ | constant in the rate of initiation of apoptosis |
| $G$ | rate of apoptosis |
| $L_{i=1,2,\ldots}$ | rate of necrosis |
| $\ell$ | constant in the expression for the rate of necrosis |

We assume that all viral activity, such as replication and new virion assembly is ceased in dead cells ($D$, $N$) and also in the cells that are undergoing apoptosis ($A$). As far as infected cells go, we have to identify several "degree of infection" measures for modelling purposes, to set up rates of various processes. When the count of the viral genomes in a cell, $i$, reaches an order of magnitude of a large enough value, $n$, then the rate of necrosis will become significant. We took $n = 10000$, selected as a number significantly exceeding the estimates 5000, see Ref. 25, and 600, see Ref. 26, in the two articles that evaluate the average viral genome counts in heavily infected cells. We then took half of this value, $m = 5000$, as a representative value for a cell to be in a "well-infected steady state" in which it is still not too likely to become necrotic. Once the count of the viral genomes in this cell, $i$, reaches the order of magnitude of $m$, we expect the cell to be in the state of steadily generating viruses, most of which are released by the cell into the media. We also define the parameter $k$, initially taken $k = 100$, estimating the number of viral genomes necessary to bring about an active infection, which would result in cell machinery being effectively "hijacked" to produce viruses.

The studied data[15] were collected in a regime of low-confluency of cell culture, though these cells are expected to divide. We fitted the rate, $R$, of heathy-cell division from an experiment for different cells (canine instead of chicken), which were carried out by us independently (not reported here). The result is conveniently written as $\ln(2)/R = 27.0 \pm 0.4$ hours. The cell-division rate of doubling in about 24 hours is typical[27] for eukaryotic cells. The rate equation for the concentration of the healthy (not infected) cells, $C_0$, will then have the term with this rate:

$$\frac{dC_0(t)}{dt} = RC_0(t) - I_0(t)C_0(t) \ . \tag{3}$$

The second term in this rate equation corresponds to the rate of viral infection, described below. Note that, although it is known that infected cells with a small number of viral genomes in them can still divide, we considered this process negligible in the modelling of the present system.

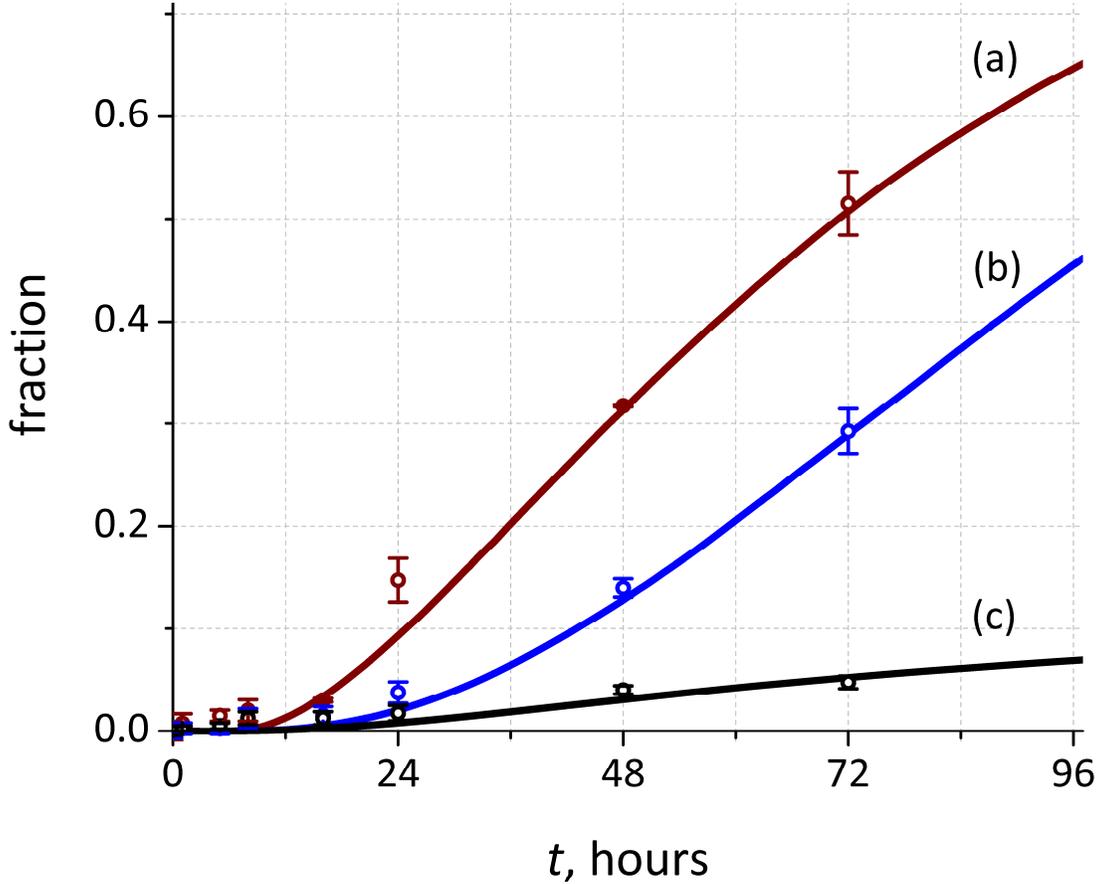

**Figure 1.** Fit, solid lines, of the available experimental data,[15] the latter shown as open circles with the corresponding error bars. Fractions $(A + D)/S$, $(D + N)/S$, and $N/S$ are marked as (a), (b) and (c), respectively. The parameter values used here are given in Sec. 3.

The concentration of the viruses in the culture, outside of the cells, $V(t)$, varies in time as viruses enter and exit the cells. We assume that the cell culture remains sufficiently low-confluency that the change in their number does not affect the outside volume to require accounting for by recalculating various concentrations. Furthermore, since the data are all for the various dimensionless ratios (fractions) of the cells' counts or concentrations, the precise definition of these quantities and their units is irrelevant here. Initially all of the cells in the system are non-infected. With these definitions, we introduce the rate, $I_{i=0,1,2,...}(t)$, for the process of endocytosis,[28] i.e., penetration of viruses into a viable non-infected or infected (but not apoptotic or dead) cell. Note that $I_0$ enters in Eq. (3).

Generally, the rate of infection $I_{i=0,1,2,...}$ should depend on the concentration of viruses outside, $V(t)$, which means that this rate itself is time-dependent. It should also depend on the degree of infection of the cell, $i$. We took the following form:

$$I_i(t) = r \frac{V(t)}{V(t) + mC_0(0)} \times \frac{m}{i+m}. \tag{4}$$

Here $r$ is the overall rate constant, fitted later from the data. The dependence on $V(t)$ should be linear for short times, when there are few viruses outside the cells, because the intake will be rate-limited by the diffusional transport of viruses to the cells. However, as $V(t)$ increases, the cells will no longer be able to accept most of the viral particles landing in its surface because there is a limit to the uptake capability of the cell membranes. We phenomenologically modelled this by using the rational expression $V/(V + mC_0)$, so that infection rate saturates as infection progresses. We took the saturation factor as $m$ in order not to introduce another large-value parameter in the model. In addition, we assume that virus intake by the "well-infected" cells' membranes will be impeded. The nature of this limit lies in the limited supply of cell surface receptors, necessary for virus docking and endocytosis. Additionally, since exocytosis and endocytosis rely on the same factors, cells that rapidly produce and release new viral particles will be partially resistant to new virus entry. Since this process fully develops at infection level $m$, we introduced the factor $m/(i + m)$, to suppress the intake for $i \gg m$.

The infection rate just considered enters the rate equation for the concentration of the infected cells,

$$\frac{dC_{i=1,2,...}}{dt} = (I_{i-1} + P_{i-1})C_{i-1} - (I_i + P_i)C_i + B_{i+1}C_{i+1} - B_iC_i - Q_iC_i - L_iC_i. \tag{5}$$

Note that not only $C_i(t)$ but also $I_i(t)$ are time-dependent, see Eq. (5). All the new terms entering Eq. (5) as well as some other rate parameters are described in the rest of this section.

To describe the state at which cell machinery of the infected cell is hijacked to replicate viral genomes, we introduce the rate of viral genome production inside the infected cells, $P_{i=1,2,...}$. As before, Eq. (4), we assume a simple rational expression

$$P_i = p \frac{i}{i+k} . \tag{6}$$

The fitting results for the rate parameter $p$ are presented in Sec. 3; recall that $k = 100$ estimates the count for which the cell has its transcription processes fully hijacked.

In addition to the production of viruses, we also have to consider the process of a viral genome exiting the cell as a virus. $B_{i=1,2,...}$ represents the rate at which viruses exit infected cells. We assume that this rate first increases linearly with the number of genomes, but saturates past the count $k = 100$ of them, and, as before, we take a simple rational expression,

$$B_i = b \frac{i-1}{i+k} . \tag{7}$$

It is expected[29] that $B_k \simeq 1200 \text{ hour}^{-1}$. Thus, we set $b = 2400 \text{ hour}^{-1}$ in order to obtain this value of $B_k$. The terms with these rates appear in Eq. (5). The last two terms, representing respectively the processes of entering apoptosis and necrosis, will be explained shortly. The process of viruses entering and exiting cells affect the concentration of viruses outside, $V(t)$, according to the rate equation

$$\frac{dV(t)}{dt} = \sum_{i=1,2,...} B_i C_i(t) - \sum_{i=0,1,2,...} I_i(t) C_i(t) . \tag{8}$$

Infected cells may enter apoptosis.[30-33] We use $A(t)$ to represent the concentration of apoptotic cells. Since, during apoptosis a number of endonucleases are activated to "shred" all internal DNA,[31] we assume that all the viral activity in apoptotic cells is effectively ceased. The parameter $Q_{i=1,2,...}$ is the rate at which infected cells respond to the infection by entering apoptosis. We assume the simplest increasing, but saturating, rational function,

$$Q_i = q \frac{i}{i+m} , \tag{9}$$

where $q$ is fitted from the data (Sec. 3), and the saturation occurs past $m = 5000$, introduced earlier as a representative value for a cell to be in a "well-infected steady state." The rate equation for the concentration of apoptotic cells is

$$\frac{dA}{dt} = \sum_{i=1,2,\ldots} Q_i C_i - GA . \tag{10}$$

Here the rate at which apoptotic cells die, $G$, is fitted from the data. The concentration of cells that died via apoptosis is denoted $D(t)$,

$$\frac{dD}{dt} = GA . \tag{11}$$

Cells can also die by necrosis. Necrosis may occur due to the depletion of cellular resources or due to the physical damage from infection, such as a cell rupturing from a large quantity of viruses being released.[32,34,35] We do not consider the latter process. $N(t)$ represents the concentration of dead cells due to necrosis (necrotic cells). We assume that when a cell enters necrosis, all viral activity in it ceases and it can be counted as practically immediately dead. $L_{i=1,2,\ldots}$ is the rate for the process of a viable-infected cell to die by necrosis. We take this rate as

$$L_i = \ell \frac{i+k}{i+n} , \tag{12}$$

where $n = 10000$ and $k$ were introduced earlier in this section, and $\ell$ is fitted. The concentration of necrotic (dead) cells is described by

$$\frac{dN}{dt} = \sum_{i=1,2,\ldots} L_i C_i . \tag{13}$$

To recapitulate, the dynamics of the process is modelled by the set of rate equations Eqs. (3), (5), (8), (10), (11) and (13), which are solved numerically with the initial condition that all the cell-type concentrations are zero, except $C_0(0)$, and with $V(0) = C_0(0)$ for the virus concentration in the considered experiment.[15] Since we are only interested in certain fractions, as described earlier, the actual value of $C_0(0)$ is immaterial for the modelling purposes. As mentioned earlier, a conventional numerical solution of the set of rate equations of the type introduced in this section is numerically costly.

Therefore, for faster simulation and parameter fitting we implemented a variation of one of the Stochastic Simulation Algorithms (SSA). These have been developed[36-40] for chemical kinetics and various biological processes and are available in different software packages, e.g., StochKit2, see Ref. 41. Originally, the Direct Method was introduced in Ref. 42 and recently optimized for a large number of applications. In the Appendix, we describe the variant — a stochastic ensemble approach — that we found particularly suitable for our problem. It differs from the standard approach, designed primarily for chemical reactions[39] by the choice of the next reaction event and the time when to proceed with that reaction. Furthermore, our approach results in a self-averaging system, thus requiring statistical averaging over only a few independent runs. Technical details are described in the Appendix.

## 3. RESULTS AND DISCUSSION

The results of data fitting for the system explored by Kumar et al.,[15] are shown in Fig. 1. Later in this section we discuss which of the system parameters can be reliably extracted from the properties for which experimental data are available. One of the advantages of theoretical modelling is the ability to quantify and explore hidden features of the modelled systems. Those might not be directly measurable but available through numerical simulations. We illustrate how the model parameters affect the quantities that were not probed in the experiment.

Figure 2 addresses the role of cell division and the fact that it is only significant during an initial period of time, for approximately seven hours, by which time most of the cells get infected, and then assumed (in our model) not to replicate. Specifically, Fig. 2A shows how the total number of cells, $S(t)$, varies with time relative to its initial value, $S(0) = C_0(0)$. The total number of cells levels out after the initial increase for about seven hours. The growth stops because most cells are by then infected, as illustrated in Fig. 2B. The latter observation is in agreement with that the average infection time is of the order of three hours.[43] The analyzed data[15] are also consistent with that, for times of up to approximately seven hours, all of the cell types the production of which depends on apoptosis and necrosis were measured to be practically zero (Fig. 1).

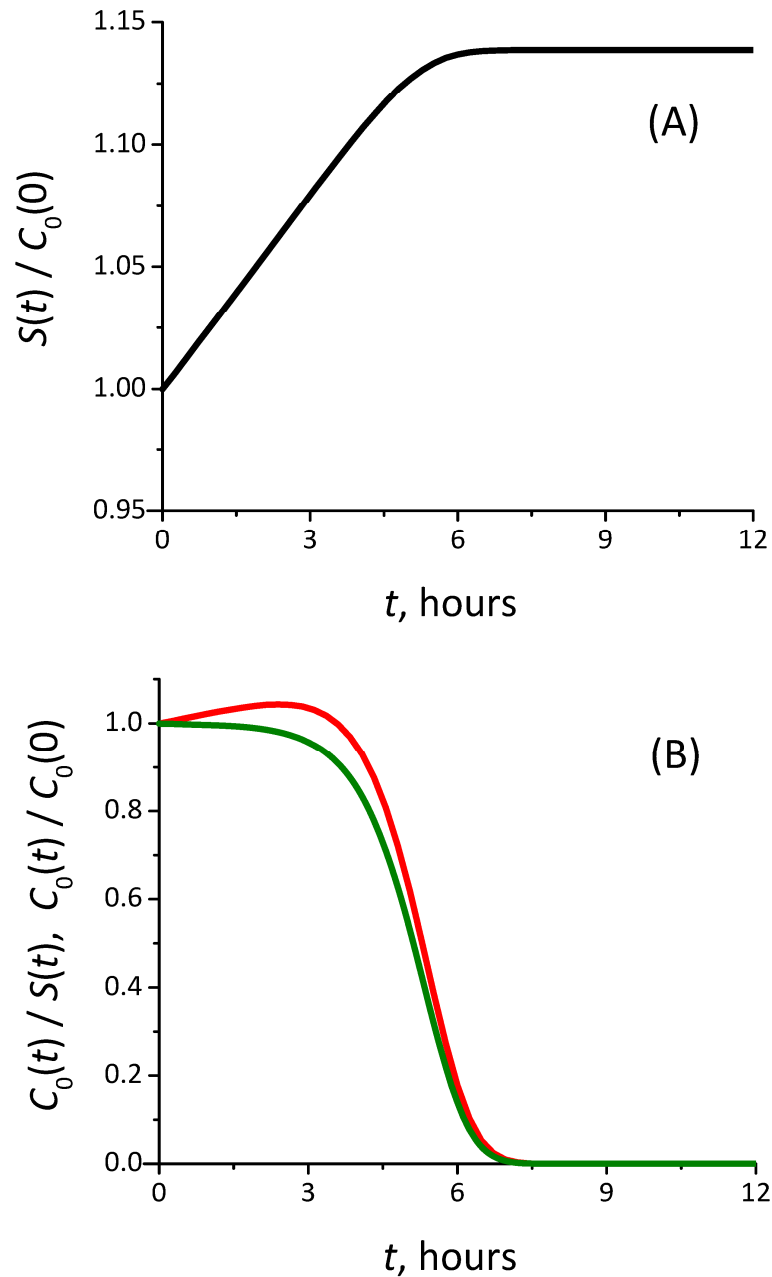

**Figure 2.** (A) The time-dependent total concentration of all cells divided by their initial concentration, $S(t)/C_0(0)$. (B) Time dependence of the fraction of healthy non-infected cells, $C_0(t)/S(t)$, monotonically decreasing due to infection (green curve). Also shown (red curve) in the ratio $C_0(t)/C_0(0)$.

**Table 2.** Values of model parameters, summarized in the Nomenclature Chart (Table 1) and in the text that were either fitted or taken from related experiments, and description of the quality of their determination when applicable.

| Parameter value | Description |
| --- | --- |
| $n = 10000$ | Taken to significantly exceed the values from Refs. 25, 26, where experiments were carried out for different types of cells. The analyzed data are not sensitive to this value. |
| $m = 5000$ | Taken as approximately $n/2$. |
| $k = 100$ | Chosen to yield the correct time scale of the onset of infection and is correlated with the parameter $r$ that was fitted afterwards. The data[15] are more sensitive to the choice of $k$ than to $r$, but not sufficient to determine this parameter precisely; this quantity is expected to be specific to certain types of viruses. |
| $r = 15.25 \text{ hour}^{-1}$ | Fitted, but depends on the assumed value of $k$. The fitted value results in the expected infection onset time. |
| $b = 2400 \text{ hour}^{-1}$ | Taken from results by Timm and Yin.[29] |
| $p = 2650 \text{ hour}^{-1}$ | Fitted precisely, given the value of $b$. |
| $\ell = 0.0029 \text{ hour}^{-1}$ | Fitted from the data, but correlated with the parameters, $b$ and $p$, which affect the degree of infection as a function of time. |
| $q = 0.0203 \text{ hour}^{-1}$ | Determined by the slope of the apoptotic cell fraction curve, given the values of $b$ and $p$. |
| $G = 0.0231 \text{ hour}^{-1}$ | Determined directly by fitting the fraction of apoptotic cells. |
| $R = 0.0257 \text{ hour}^{-1}$ | The analyzed data are not sensitive to this parameter, and therefore its value was taken from another experiment carried out by us. Large values of parameter $r$ make our model practically unresponsive to this parameter. |

Model parameters that are the most relevant for the regime just considered, Fig. 2, are $R$, $r$, $b$, $p$, and $k$. Their fitting or determination from other data were commented on earlier, in Sec. 2, and are summarized in Table 2. Some of these parameters control the kinetics of the processes at later times, but some, such as $R$, do not affect the later-time behavior and their values cannot be precisely fitted from the considered data. Generally, see Table 2, several parameters are well-determined by the analyzed data of Ref. 15, but several other parameters are not, and we took these from other sources (Sec. 2).

Once the infection is well-developed, the model allows us to consider quantities related to its progress. Figure 3 illustrates the time dependence of the count of infected cells, $\sum_{i=1,2,\dots} C_i(t)$, normalized by the total number of cells, $S(t)$. The sharp feature in the data at time 7 hours is magnified to demonstrate that the process happens smoothly, despite the fast onset of the effects of infection.

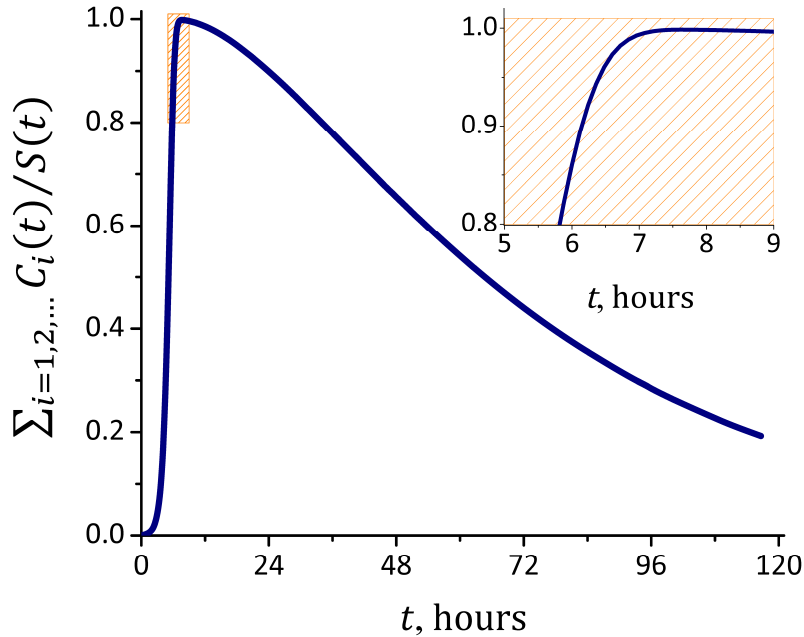

**Figure 3.** The fraction of the infected cells that did not yet enter apoptosis or died by necrosis. The Inset demonstrates that the behavior near the peak is smooth.

Additional calculated rather than measured quantities of interest include the number of viruses in the culture and the average number of viral genomes (the degree of infection) in cells that are not apoptotic or dead, see Fig. 4. An interesting observation is that the number of viral genomes in infected (but not apoptotic or dead) cells increases approximately linearly, at the rate $p - b$, for the considered time scales, Fig. 4B, after the initial 7-hour interval identified earlier.

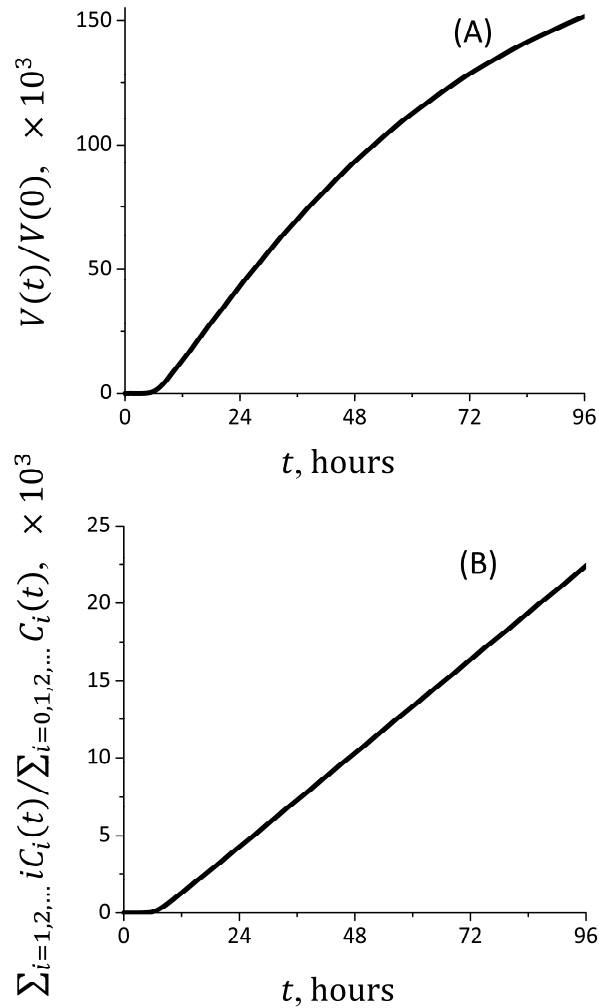

**Figure 4.** (A) Number of viruses in the culture, normalized to the initial total number of viruses (and shown in thousands). For larger times, this curve will ultimately saturate due to necrosis and apoptosis. (B) The average number (shown in thousands) of viral genomes (the degree of infection) inside cells that did not yet enter apoptosis or died by necrosis.

To verify and validate the utility of our model, we set to apply our calculations to the experimentally measured dynamics of infectious bursal disease virus (IBDV) in cell culture. In the work presented by Rekha et. al.,[44] the authors infect cell culture of chicken embryo fibroblasts with IBDV and measure virus titer every 12 hours using TCID50 (tissue culture infectious dose) assay. The progression of the infection (time-dependent increase of the virus titer) is presented in their paper in Figure 5. We have extracted the values for the virus titer during the first 60 hours of infection and tested our model on these data. The results are presented in Fig. 5.

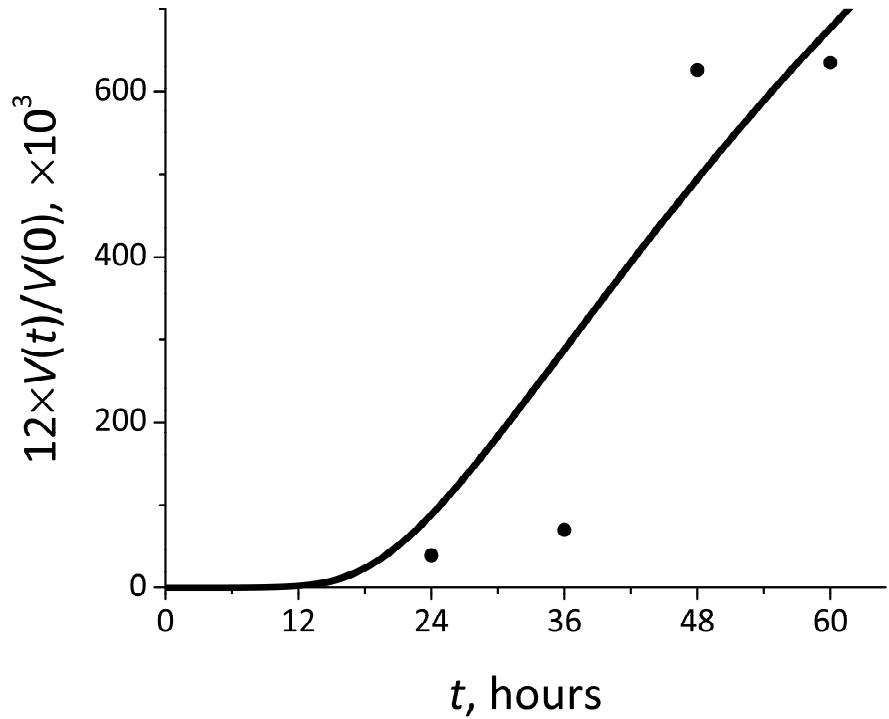

**Figure 5.** Comparison of IBDV infection progression computed with our model: solid line, and measured experimentally (Ref. 44): data points. The factor 12 (only for the solid line) was fitted to account for that not all the viruses in the experiment of Ref. 44 were active, which was not assumed in our original model. Note: at time $t = 0$ the line starts at the value 12 (not seen on the shown scale)..

TCID50 units used in experimental work had been converted to plaque-forming units (pfu) using classical relationship, and we used particle-to-pfu ratio for IBDV as 2360:1, demonstrated in Ref. 45. This number is very close to the value experimentally determined for IBDV using reverse transcription.[46] Thus for model curve in Fig. 5 we used the value of $k = 2360$, and adjusted parameters to $r = 1220$ hour$^{-1}$ and $\ell = 0.0021$ hour$^{-1}$ in order to provide a precise fit of the same quality as in Fig. 1. Note, with these parameters modified the quantities presented in Fig. 2 and Fig. 4 would be quantitatively different. For instance due to higher infection rate the time-dependent degree of infection is somewhat increased, and the number of viruses released is smaller, compared to those in Fig. 4A.

Long-term experimental data[44] (beyond 3 days of culturing) shows decline in virus titer, likely due to the degradation of both cells and viruses in the culture as a result of deteriorating culturing conditions, such as exhaustion of nutrients and increased media acidity, conditions not included on our model and therefore not presented here. In general, these data demonstrate that our model describes IBDV infection dynamics adequately and can be used to extract parameters, which are hard to measure experimentally.

To further elucidate the utility of the model we investigated the probability distribution of the number of viral genomes in live cells (the degree of infection), and the evolution of this distribution with time. This quantity is related to the burst size (number of virus particles produced per infected cell). The resulting dynamics (with the original model parameters) is presented in Fig. 6. We find that the burst size is a probabilistic value, which is approximately normally distributed and nearly linearly increasing in time during progression of the infection.

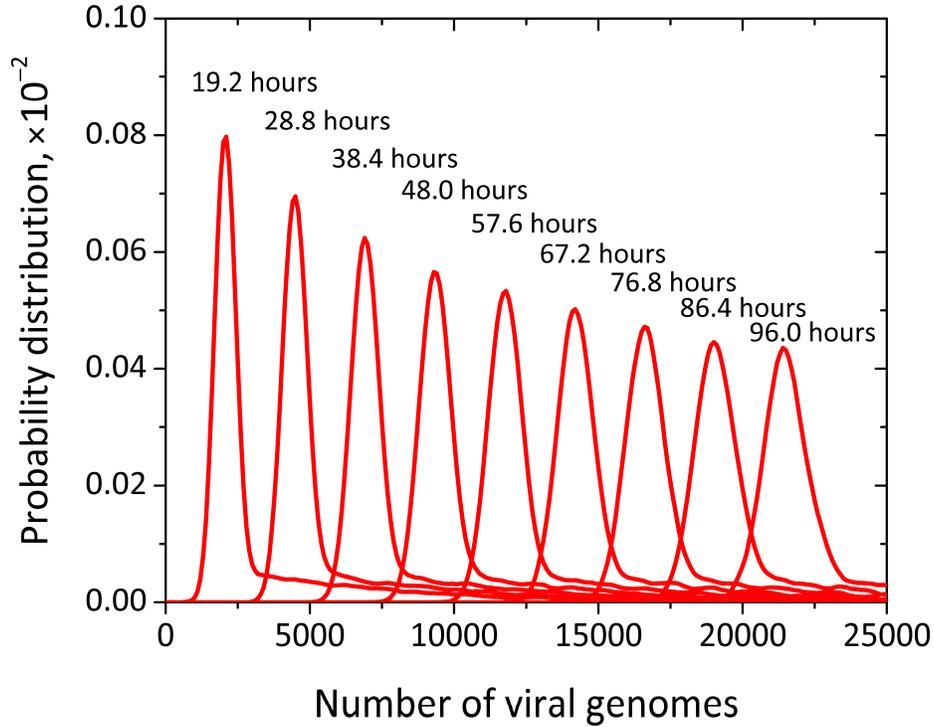

**Figure 6.** Dynamics of the probability distribution of the number of viral genomes in live (and non-apoptotic) infected cell, $C_{i=1,2,\ldots}(t)$, as a function of time, depicted multiplied by a factor that makes the curves better visible on the shown scale. Note that the total count of such cells is plotted in Fig. 3.

In summary, our results indicate that the modelling of biological processes allows understanding the cell dynamics under stress in greater details: The presented approach allows extraction of some of the model parameters, which characterize the time-dependent data that were measured, and calculation of properties that were not measured but may be of interest in understanding the system's behavior. In the case of viral infection progression, we can extract novel, unmeasured values, such as viral load per cell, or total number of viral particles in the cell culture at any given time. An important point to be made is that our model does not use an explicit particle-to-pfu ratio, which in reality is a combination of numerous phenomena, such as quality of viral particle production, threshold concentration of viral particles necessary to initiate infection cascade, etc. In lieu of particle-to-pfu ratio we utilize parameter $k$, closely related to this number. This parameter defines number of viral particles necessary to fully push an infected cell towards viral production. The modelling approach can also suggest other quantities to measure, in order to better understand the behavior of the system. The considered rate-equation modelling with the stochastic ensemble approach applies to low-confluency cell cultures, and is complementary to models that focus on the system's cluster properties.[1,2]

## ACKNOWLEDGEMENTS

The work was supported in part by a grant from American Federation for Aging Research (AFAR) 2015 to SL.

## APPENDIX. STOCHASTIC SIMULATION ALGORITM

This appendix describes the stochastic "ensemble" approach for emulation of the dynamics of our system. With high accuracy, the results are equivalent to the numerical solution of the rate equations introduced in Sec. 2. The latter were solved with adaptive-step RKF45 method,[23, 24] which for stiff equations requires decreasing the time-discretization steps to very small values in order to maintain the accuracy. The implemented variant of the SSA resembles Direct Method (DM) by Gillespie.[42] It is computationally efficient and therefore can be of interest in other situations involving similar systems. The utilized "ensemble" approach is based on stochastically evolving a large number of objects (cells) labelled with various cellular properties, e.g., healthy, infected to various degrees, undergoing processes of apoptosis and necrosis, etc. Our approach differs from the standard utilization by the choice of the next reaction event. We use a probabilistic rate-weighted choice of the next reaction, conceptually similar to Ref. 42, but correlated with the choice of the next reaction event time. In the DM, the treatment of these quantities is explicitly separated.

The major difference from the DM is the order of the reaction events—in DM the order is randomized with weighted probabilities, while in our SSA it is deterministic, i.e., during a small time interval (the "synchronization" time, $\Delta t$) event A is carried out consecutively several times (depending on $\Delta t$), then event B is carried out consecutively, then event C, etc. This allows spanning through the iterations faster, however, with the tradeoff of introducing a small systematic error because of the time- and $i$-dependence of the rate parameters.

There is also a difference in the averaging of the quantities. In the DM, every realization is propagated up to $t_{max}$, the maximum desired simulation time, then a large number of realizations is averaged over. With the present stochastic ensemble approach, all of the quantities of relevance are self-averaging, provided the ensemble is large enough, thus one or few realizations may suffice to yield low-noise data.

The stochastic ensemble simulation algorithm is outlined below:

**Step 1.** Create an ensemble: a set of objects stored as a one-dimensional array, in which each object is a cell of a certain cell-type, including the degree of infection. Initially, the array has size $C_0(0)$. We also set the variable $V = V(0)$, and calculate all the $i$-dependent factors in the rate expressions.

**Step 2.** Propagate each process consecutively by time $\Delta t \ll t_{max}$. For this, iterate each of the moves 2.1-2.7, listed shortly, as long as

$$\alpha \sum_{j=1,2,\ldots} 1/s(j) < \Delta t \tag{A1}$$

holds, where

$$\alpha = 1/\max[\Delta t^{-1}, \{X_{i=0,1,\ldots}\}] . \tag{A2}$$

Here $j$ is the current iteration index, $s(j)$ is the size of the current subset of cells relevant for the current move, such as cell division, cell in which viral genome production occurs, etc. This subset size may change with iterations, and $\{X_{i=0,1,\ldots}\}$ are the rate constants for the relevant process, defined in Sec. 2. Before the last iteration (we denote its index as $j + 1$) in each time interval, $\Delta t$, i.e., when Eq. (A1) is not satisfied, recalculate $\alpha$ according to:

$$\alpha = [\Delta t - \alpha \sum_{j=1,2,\ldots} 1/s(j)]s(j + 1) . \tag{A3}$$

- **2.1.** Pick a non-infected cell, and duplicate it with probability $\alpha R$.
- **2.2.** Pick a live and non-apoptotic cell, and add one viral genome with probability $\alpha I_i$. Decrease $V$ by one.
- **2.3.** Pick a cell with $i \geq 2$, export one virus with probability $\alpha B_i$. Increase $V$ by one.
- **2.4.** Pick an infected viable cell, add one virus with probability $\alpha P_i$.
- **2.5.** Pick an infected non-apoptotic live cell, initiate apoptosis with probability $\alpha Q_i$.
- **2.6.** Pick an apoptotic cell, make it dead by apoptosis with probability $\alpha G$.
- **2.7.** Pick a non-apoptotic viable cell, make it dead by necrosis with probability $\alpha L_i$.

**Step 3.** At convenient time intervals calculate data from the ensemble.

**Step 4.** If the time is less than $t_{\max}$, recalculate $I_i(t)$ and repeat the Steps 2-4 above. Note that this time-dependent rate does not change significantly within $\Delta t$, therefore there is no need to recalculate it during every iteration.

For optimization purposes one can create a list for each object subset, i.e., list of cells for division, list of cells for virus replication, etc. These lists store identifiers of objects from the ensemble, for fast retrieval.

**Note 1:** In moves 2.1-2.7, the objects should be picked randomly from a proper subset.

**Note 2:** Decreasing $\Delta t$ will reduce the error (as compared to the numerical solution of the rate equations), but will somewhat increase the computation time, because the quantities $\alpha X_{i=0,1,\ldots}$ may decrease, thus resulting in more rejection events and in time advancing in much smaller steps.